\documentclass[11pt]{article}
\usepackage{moriond,epsfig}
\usepackage{amsmath,amssymb,graphicx}
\usepackage{pstricks}

\def\be{\begin{equation}}
\def\ee{\end{equation}}
\def\bea{\begin{eqnarray}}
\def\eea{\end{eqnarray}}

\begin{document}
%\preprint{ANL-HEP-PR-04-19, EFI-04-06}
ANL-HEP-PR-04-62, EFI-04-22

\vspace*{4cm}
\title{PROTON STABILITY AND DARK MATTER: ARE THEY RELATED?}

\author{ G\'eraldine SERVANT }

\address{High Energy Physics Division, Argonne National Laboratory, Argonne, \\
Enrico Fermi Institute, University of Chicago, Chicago, \\
Service de Physique Th\'eorique, CEA Saclay, France}

\maketitle\abstracts{
We address the problem of baryon number violation in Randall-Sundrum backgrounds and provide a  solution leading to a stable light Kaluza--Klein fermion in warped GUT. This adds to the list of dark matter candidates which stability can follow from ensuring proton stability in weak scale extensions of the Standard Model.}

\section{Baryon number violation and dark matter candidates in weak scale extensions of the Standard Model}

In this talk, we provide a new  example of a connection between proton stability and dark matter. The best-known example of such connection arises in the Supersymmetric Extension of the Standard Model (SSM) where the stability of the dark matter candidate, namely the Lightest Supersymmetic Particle (LSP), results from imposing R-parity. R-parity is not imposed just to get a dark matter candidate but as a simple solution to the problem of baryon number violation in the SSM. 
In fact,  the lagrangian of the SSM contains {\it a priori}
 dangerous four-dimensional renormalizable operators, consistent with supersymmetry as well as all the gauge symmetries of the Standard Model, but which violate baryon (B) and lepton (L) numbers. They come from the superpotential:
\begin{equation}
W=\lambda_{ijk}L_iL_jE_k^c+\lambda'_{ijk}L_iQ_jD_k^c+\lambda''_{ijk}U_i^cD_j^cD_k^c 
%+\kappa_i L_iH_2
\end{equation}
 On the other hand, these interactions are odd under the discrete R-parity:
 \begin{equation}
 R=(-1)^{3(B-L)+2S}
 \label{Rparity}
 \end{equation}
 The simplest way to forbid B and L violation from these operators is thus to invoke R-parity conservation. As an interesting consequence, the LSP is stable and provides a well motivated WIMP dark matter (DM) candidate.
 Note that  there are other solutions to the proton stability problem in the SSM which do not lead to a stable DM particle. Those are lepton parity (allowing only the $U^cD^cD^c$ operator) 
 or baryon parity (forbidding only the $U^cD^cD^c$ operator), which maintain the proton stable while allowing the LSP to decay.
 
 The solution containing conserved R-parity is very attractive since it provides 
 an ideal WIMP DM candidate. Indeed, the freeze-out relic density calculation of 
 a particle with weak scale mass and  weak scale size interactions generically leads to precisely the right energy density contained in DM.  It only requires the weak assumption that the reheat temperature of the universe should be at least a few tens of GeV to justify that the WIMP is initially at thermal equilibrium with the rest of the primordial plasma. One then finds that it  typically decouples from the thermal bath at a temperature $T_f\sim m/x_f$ where $x_f\sim25$, with a relic energy density given by
  \begin{equation}
\Omega_{\mbox{\tiny{relic}}} h^2\sim \frac{ 10^9}{M_{Pl}}\frac{x_f}{\sqrt{g_*} }
\frac{\mbox{GeV}^{-1}}{\langle \sigma v\rangle}
\label{RELIC}
\end{equation}
If the thermal average of the annihilation cross section (times the relative velocity) is 
$\langle \sigma v\rangle \sim 10^{-9}$ GeV$^{-2}\sim 1$ pb, as given by a typical electroweak interaction, 
then $\Omega_{\mbox{\tiny{relic}}} h^2 \approx  \Omega_{\mbox{\tiny{DM}}} h^2 $.
  Thus, a weak scale mass LSP or any new neutral weak scale particle  appear
   as a very natural explanation for DM. It is also well motivated as we expect new physics at the weak scale to solve the hierarchy problem.

 That B conservation
is no more guaranteed in the SSM is a generic difficulty arising in most extensions of the Standard Model (SM).  Little Higgs theories
may also resolve the problem by implementing the discrete symmetry (\ref{Rparity}). These theories
comprise new degrees of freedom at the weak scale which can be odd under $R$ and  play the role of DM \cite{Katz:2003sn}.
 
In the last five years, non supersymmetric but extra-dimensional approaches to the hierarchy problem have been proposed. The one which attracted most attention being the Randall--Sundrum (RS) model \cite{Randall:1999ee} which offers an alternative framework for understanding EW symmetry breaking, hierarchy in fermion masses, gauge coupling unification etc.
So far, a dark matter candidate was missing in these constructions. In the following, we present a model where a light Kaluza--Klein fermion is stable as a consequence of imposing  baryon number symmetry.
In the SSM, there is no theoretical reason to prefer models with conserved 
R-parity versus those with R-parity violation. However, models with conserved R-parity possess a nice WIMP dark matter candidate for free.
Similarly, in warped geometry, as we will show, ensuring proton stability does not necessarily results in a stable exotic particle. 
We choose here to study the solution leading to it and to investigate the properties of this dark matter candidate. In the present article, we only sketch the main ideas. Details can be found in \cite{Agashe:2004ci,RSLKP2}.
\section{An alternative to the Supersymmetric SM: the SM in warped space}
The RS geometry for solving the hierarchy problem consists of a compact slice of $AdS_5$ bounded by two branes: the $UV$ (Planck) and IR (TeV) branes. The TeV scale on the IR brane is generated from the Planck scale via the
exponential warped factor of the $AdS_5$ metric. 
In the original RS set-up, the SM was localized on the IR brane and only gravity could propagate in the 5-dimensional anti-de-sitter space. 
%There is a 4d massless graviton which wave function is peaked on the Planck brane. 
The effective cut-off scale on the IR brane  being TeV, if the Higgs is localized on that brane, the hierarchy between the Higgs mass and the Planck scale is protected. On the other hand, there is no need for localizing SM fermions and gauge fields on the TeV brane. It is actually more favourable to delocalize the SM fields into the bulk, as it allows for possible unification of gauge couplings \cite{running} as well as a mechanism to generate the hierarchical spectrum of Yukawa couplings 
\cite{neubert}. The most promising models to date are those extending the EW gauge group to $SU(2)_R\times SU(2)_L\times U(1)$. The additional $SU(2)_R$ gauge factor in the bulk is required to reproduce the global custodial symmetry of the SM \cite{Agashe:2003zs}.
\begin{figure}[h]
\begin{center}
%\rule{5cm}{0.2mm}\hfill\rule{5cm}{0.2mm}
%\vskip 2.5cm
%\rule{5cm}{0.2mm}\hfill\rule{5cm}{0.2mm}
\psfig{figure=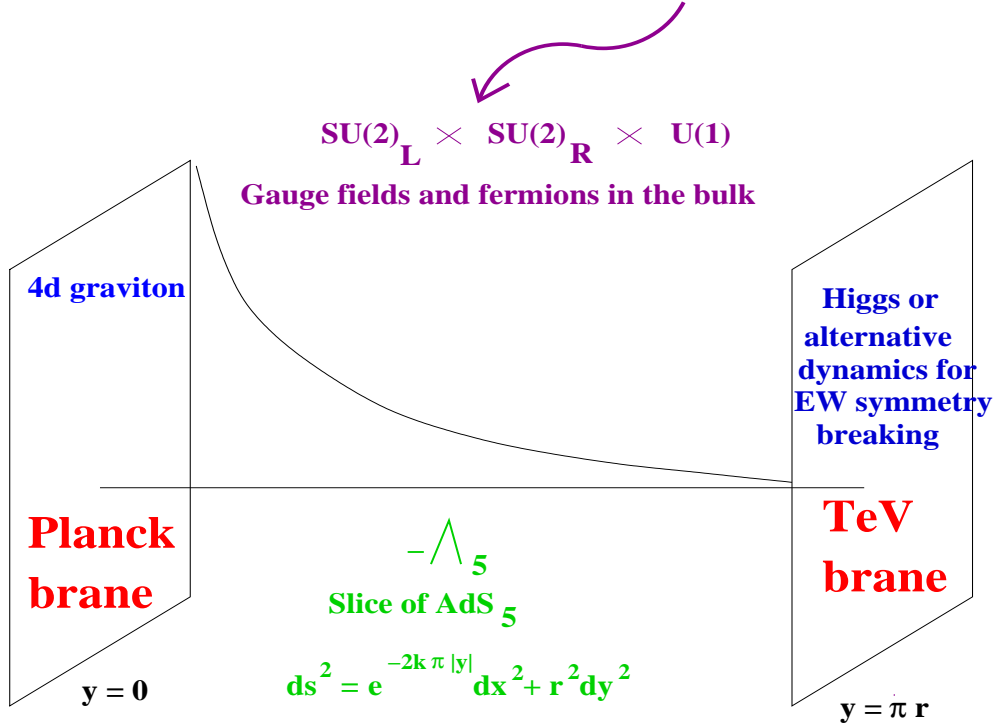,height=3.8in,width=5.2in}
\caption{A realistic model: SM gauge fields and fermions are bulk fields. Only the Higgs 
 or alternative dynamics for electroweak symmetry breaking is peaked on the IR brane. 
In addition, the EW gauge group has to be extended to $SU(2)_L\times SU(2)_R \times U(1)$ to guarantee the existence of a global custodial symmetry in the 4D CFT. 
Therefore, when considering a GUT in warped space, $SO(10)$ will be preferred to $SU(5)$.
4D Yukawa couplings are given by the overlap of the wave functions of the fermions with the Higgs.  Consequently, only the top quark is expected to be near the TeV brane while other SM fermions are closer to the Planck brane. 
\label{fig:framework}}
\end{center}
\end{figure}

\subsection{Baryon number violation in warped space}

 The dangerous baryon number violating interactions come from effective 4-fermion operators, which, after dimensional reduction lead to: 
\begin{eqnarray}
\int dy \ d^4x \ \sqrt{-g} \ \frac{\overline{\Psi}_i\Psi_j\overline{\Psi}_k\Psi_l}{M_5^3}\sim 
\int d^4x \ e^{\pi k r_c(4-c_i-c_j-c_k-c_l)} \ \frac{ {\overline{\psi}}_i^0 \psi_j^0
 {\overline{\psi}}_k^0 \psi_l^0}{m_{Pl}^2} 
 \label{dangerous}
\end{eqnarray}
where $i,j,k,l$ are flavor indices. The $\psi^0 $ are the 4D zero mode fermions identified with the SM fermions. The localization of their wave functions along the fifth dimension is parametrized by the $c$'s, which therefore control the size of the 4D Yukawa couplings with the Higgs.
To obtain a Planck scale suppression of this operator,  the $c$'s have to be larger than 1, meaning that zero mode fermions should be very close to the UV brane.
Unfortunately, this is incompatible with the Yukawa structure,
 which requires that all $c$'s be smaller than 1. 
 
 When working in a GUT, there is an additional potential problem coming from the
exchange of TeV KK excitations of grand unified gauge bosons, such as X/Y gauge bosons, 
mediating fast proton decay.
Indeed, it turns out that all TeV KK modes and therefore X/Y TeV KK gauge bosons are 
localized near the TeV brane. Their 
 interactions with zero mode fermions will be suppressed only if fermions are localized very close to the Planck brane, again requiring that $c$'s be larger than 1. 
This problem arises in any GUT 
theory where the X/Y's propagate in extra dimensions with size larger than $m_{GUT}^{-1}$.
A simple solution to this problem 
%suggested by  \cite{}
 is to assume that SM fermions 
come from different GUT multiplets. Concretely, this means that boundary conditions are not the same for all components of a given GUT multiplet so that only part of the fields in a multiplet acquire zero modes, which are identified with SM particles. 
While this circumvents the problem of B violation due to KK X/Y exchange, one still has to cure the B violation due to the effective operator (\ref{dangerous}). We do so by imposing an additional symmetry. 

\section{Imposing gauged baryon number symmetry in warped GUT}
We illustrate the idea with a simple example  where $SO(10)$\footnote{The motivation for $SO(10)$ is that it contains the precious $SU(2)_R$ while $SU(5)$ does not.} is broken to the SM 
on the Planck brane by boundary conditions
 so that SM quarks and leptons are obtained from different bulk multiplets of the unified gauge 
  group. 
 % To repeat,
%such procedure forbids  proton decay  coming from the s-channel exchange of TeV KK excitations of $X/Y gauge bosons.
Besides, the number of $\bf{16}$ representations has to be replicated at least
three times per generation:
$$
  \begin{pmatrix}\bf{u_L \ , d_L}\\u^{\prime c}_R \\ d_R^{\prime c}  \\ 
  \nu^{\prime}_L \ ,  e^{\prime}_L \\ e_R^{\prime c} \\ \nu_R^{\prime c} \\\end{pmatrix}_{B=1/3}, \ \ 
 \begin{pmatrix}u^{\prime}_L \ , d^{\prime}_L\\ \bf{u_R^c} \\ \bf{ d_R^c}  
 \\ \nu^{\prime}_L \ ,  e^{\prime}_L \\ e_R^{\prime c} \\ \nu_R^{\prime c} \\\end{pmatrix}_{-1/3}, \ \
  \begin{pmatrix}u^{\prime}_L \ , d^{\prime}_L\\u_R^{\prime c} \\ d_R^{\prime c} 
   \\ \bf{\nu_L} \ ,  \bf{e_L} \\ \bf{e_R^c} \\ \bf{\nu_R^c} \\\end{pmatrix}_{0}
$$
Only states in boldface characters have zero modes and are identified with the SM fermions.
The KK towers of the other states do not contain zero modes.
GUT multiplets are assigned the baryon number of the zero modes contained in them.
A $Z_3$ symmetry follows from requiring baryon number as a good quantum number:
\begin{equation}
\Phi \rightarrow e^{ 2 \pi i \left( B  - \frac{n_c - \bar{n}_c }{3} \right) } \Phi
\end{equation}
$B$ is baryon-number of $\Phi$ and  $n_c$ ($\bar{n}_c$)
is its number of color (anti-color) indices.
%$Z_3$ can be seen as a remnant discrete symmetry after the 
%unified gauge symmetry has been broken to the SM.
 Clearly, SM fields are not charged under $Z_3$.
$X$, $Y$, $X^{\prime}$, $Y^{\prime}$ and $X_s$ gauge bosons of $SO(10)$ are charged under $Z_3$
as well as lepton-like states within $\bf{16}$'s which carry baryon number and quark-like states 
 which carry non standard baryon number.
These exotic states do not have zero modes. 
Consequently, the lightest $Z_3$ charged particle 
 (LZP) cannot decay into SM particles and is stable.
 
B  is  broken on the Planck brane to prevent the existence of
an extra massless gauge boson in the theory. One can check that
 even if $Z_3$ is broken on the Planck brane,  the lifetime of the LZP
is $10^{10}$ times the age of the universe \cite{RSLKP2}.
 Note that an alternative solution to protect proton stability
 is to gauge lepton number in the bulk instead and break it on the Planck brane.
 Proton is still long-lived since it has to decay into the electron and this has to 
 happen on the Planck brane.
 Neutron-antineutron oscillations are allowed but still sufficiently suppressed.
 In this case, there is also an interesting phenomenology but there 
 is no stable KK particle resulting from imposing lepton number. We therefore focus on the 
 solution with baryon number symmetry.

\section{ A Kaluza--Klein right-handed neutrino as a WIMP}
\begin{figure}[h]
\begin{center}
%\rule{5cm}{0.2mm}\hfill\rule{5cm}{0.2mm}
%\vskip 2.5cm
%\rule{5cm}{0.2mm}\hfill\rule{5cm}{0.2mm}
\psfig{figure=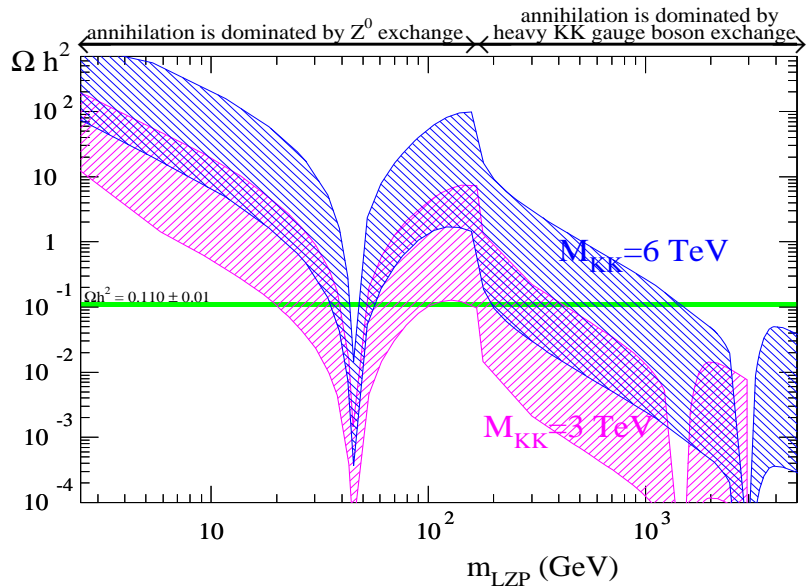,height=3.5in,width=5in}
\caption{Relic energy density of the LZP for two choices of KK gauge boson masses. Each region is obtained by varying the parameters of the model.
\label{fig:relic}}
\end{center}
\end{figure}

A firm prediction of these constructions is that the Lightest KK particle will be contained in the GUT multiplet whose zero mode is the right-handed top quark. Indeed, the mass of the first KK excitation of the GUT partners of $t_R$  depends very sensitively on the localization of the wave function of the zero mode in this multiplet \cite{Agashe:2004ci} and these KK modes become lighter as the zero mode of $t_R$ gets closer to the TeV brane, which is required to obtain the large Yukawa coupling of the top quark. There is enough freedom in our model so that the mass of the LZP can vary between a few GeV to a few TeV.
For dark matter, the only viable possibility is to identify the LZP with 
the KK right-handed neutrino \cite{Agashe:2004ci}. In warped space, such particle has large gauge couplings to  TeV mass KK gauge bosons of $SO(10)$. This results in a total annihilation cross section which can be of the right order for a large range of LZP masses as shown in fig~\ref{fig:relic}. 

\begin{figure}[h]
\begin{center}
%\rule{5cm}{0.2mm}\hfill\rule{5cm}{0.2mm}
%\vskip 2.5cm
%\rule{5cm}{0.2mm}\hfill\rule{5cm}{0.2mm}
\psfig{figure=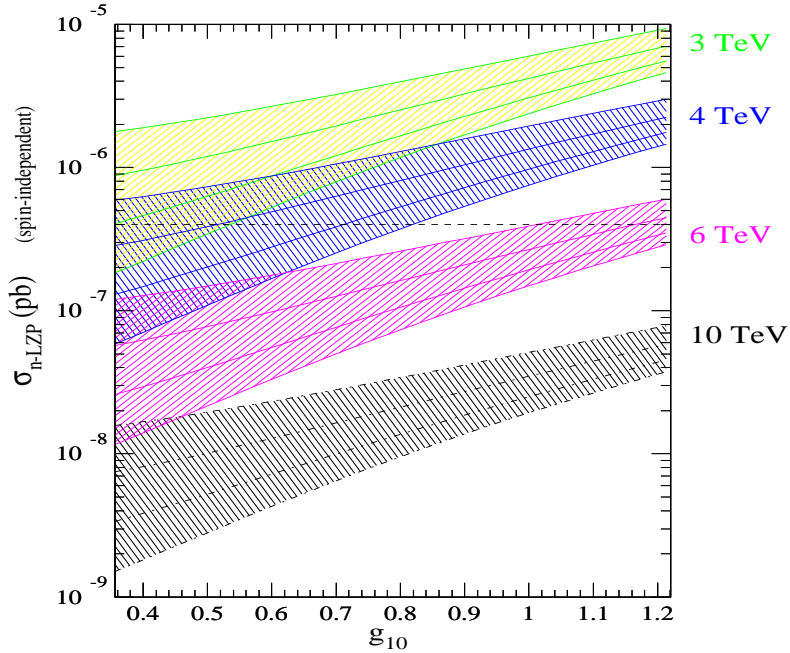,height=4.in,width=4.2in}
\caption{Predictions for elastic scattering cross section of the
 LZP off a nucleon (independent of LZP mass) for four choices of KK gauge boson masses.
 Each region is obtained by varying 
 the $Z^0$-LZP coupling.
Horizontal dotted line indicates present experimental limit, which only applies for some range of WIMP masses.
$g_{10}$ is the 4D $SO(10)$ gauge coupling which is almost a free parameter even after requiring gauge coupling unification.
\label{fig:elastic}}
\end{center}
\end{figure}

Finally, there are promising predictions for the detection of such particle through its elastic scattering with nuclei. The expected wimp-nucleon cross section is displayed in fig~\ref{fig:elastic}. It is actually large and 
some part of the parameter space can already be excluded using the present experimental constraints.
%\begin{figure}[h]
%\begin{center}
%\rule{5cm}{0.2mm}\hfill\rule{5cm}{0.2mm}
%\vskip 2.5cm
%\rule{5cm}{0.2mm}\hfill\rule{5cm}{0.2mm}
%\psfig{figure=collider_m.eps,height=2in,width=3.2in}
%\caption{
%\label{fig:collider}}
%\end{center}
%\end{figure}

\section{Summary}
$\bullet$
We have extended the possibility of Kaluza--Klein dark matter, so far restricted to models with flat and universal extra dimensions \cite{Servant:2002aq}. As far as cosmology is concerned, an advantage of the warped geometry over the flat one is that it surely does not suffer from a potential moduli problem (overclosure of the universe by radion oscillations \cite{Kolb:2003mm}).\\
$\bullet$ The motivation for the stability of the DM particle is analogous to that of the LSP in supersymmetry. 
It follows from a particular choice of solution to the proton stability problem.\\
Imposing baryon number symmetry in a warped GUT requires a replication of fundamental representations, leading to the existence of ``baryonic" leptons, the lightest of which is stable.\\
$\bullet$ Warped $SO(10)$ (and Pati--Salam) are GUT embeddings of the $SU(2)_L\times SU(2)_R\times U(1)$ models of EW symmetry breaking in warped space (with custodial symmetry) such as  \cite{Agashe:2003zs,Csaki:2003zu}. Our qualitative results apply independently of the precise nature of EW symmetry breaking provided that it happens on the TeV brane.
 In this context, the KK right-handed neutrino behaves as a typical WIMP since it has gauge interactions with TeV mass KK gauge bosons.\\
$\bullet$ The phenomenology of these models is rich. The existence of light KK fermions, which is a direct consequence of the heaviness of the top quark and a firm prediction of these models, opens the possibility
of observing Kaluza--Klein states at colliders, with distinctive signatures \cite{RSLKP2}.\\
$\bullet$ The prospects for direct detection of our DM particle with elastic scattering experiments are excellent and the entire parameter space of the model will be tested in the near future.

\section*{Acknowledgments}
Thank you to the organizers for making these {\it rencontres} very enjoyable, and  in particular
for giving me the opportunity to reconnect a bit with the french HEP community...
I thank the NSF for partial financial support of my first Moriond conference.

\end{document}